# Impact of Size and Thermal Gradient on Supercooling of Phase Change Materials for Thermal Energy Storage

Drew Lilley[1,2], Jonathan Lau[1], Chris Dames[1,2*], Sumanjeet Kaur[1*], Ravi Prasher[1,2,*]

[1]Energy Technologies Area, Lawrence Berkeley National Laboratory, Berkeley, CA 94720, USA
[2]Department of Mechanical Engineering, University of California, Berkeley, CA 94720, USA
* e-mail: cdames@berkeley.edu, skaur1@lbl.gov, rsprasher@lbl.gov

**Abstract**
Phase change material based thermal energy storage has many current and potential applications in the heating and cooling of buildings, battery and electronics thermal management, thermal textiles, and dry cooling of power plants. However, connecting lab scale thermal data obtained on DSC to the performance of large-scale practical systems has been a major challenge primarily due to the dependence of supercooling on the size and temperature gradient of the system. In this work we show how a phase change material's supercooling behavior can be characterized experimentally using common lab scale thermal analysis techniques. We then develop a statistics based theoretical model that uses the lab-scale data on small samples to quantitatively predict the supercooling performance for a general thermal energy storage application of any size with temperature gradients. Finally, we validate the modeling methodology by comparing to experimental results for solid-solid phase change in neopentyl glycol, which shows how the model successfully predicts the changes in supercooling temperature across a large range of cooling rates (2 orders of magnitude) and volumes (3 orders of magnitude). By accounting for thermal gradients, the model avoids ~2x error incurred by lumped approximations.

**Introduction**

Roughly 90% of the world's primary energy generation results in thermal energy [1,2]. The global thermal energy storage (TES) market is projected to deploy 13 GW in 2024, corresponding to a total value of $55 billion [3]. In many applications, TES has an inherent advantage over electrochemical and mechanical energy storage because its raw materials are cheap, making TES much more cost competitive in applications ranging from buildings and residential heating, to power electronics and agriculture [4]. Phase change materials (PCMs) are also being woven into thermal textiles to provide localized and personalized cooling/heating to reduce thermal load in buildings [5]. Recently PCM has also found applications in dry cooling of power plants to conserve water by utilizing the diurnal swing of ambient temperature [6]. PCM is also being actively researched for thermal management of batteries [7]. Although cost-competitive, PCM implementation is lagging relative to the abundance of thermal energy, application and opportunity.

The lag is in part due to economics, but in larger part is caused by technological barriers. For phase change materials – which this paper focuses on – those technological barriers are widely identified [4,8,9] as (1) low PCM thermal conductivity and (2) excessive supercooling, i.e. the undesirable difference between a material's thermodynamic equilibrium phase change temperature and the actual value seen in a cooling process with non-zero cooling rate. Low thermal conductivity has been addressed recently by many novel and promising techniques such as the addition of nanoparticles and the impregnation of PCM into graphite matrices [10]. However, the approaches to reduce supercooling by using either nucleating or thickening agents have met with limited success.

In a broader context, fundamental understanding of supercooling – or metastability – remains a grand challenge in the sciences. Phenomenological models, such as classical nucleation theory [10] and its various extensions, provide excellent physical insight into the nucleation process. Although physically successful, the inputs to such models (e.g surface energies, surface shape, free energy barriers) are difficult to know a-priori. For this reason, classical nucleation theory has traditionally been limited to applications requiring experimental fitting, as opposed to experimental prediction [11].

Supercooling changes with just about every material and system parameter: geometry, volume, material, microstructure, purity, discharge rate, etc. Due to the fickle nature of nucleation, predicting the performance of any large-scale practical system based on lab scale data on small scale samples is a very difficult task. Phase change is stochastic in nature, and there often exist numerous nucleation pathways that yield the same outcome [12]. As a result, when researchers report a supercooling temperature from lab-scale experiments, that temperature means little beyond that specific experimental system, size, material, and environment. This hamstrings system-level modeling of PCMs in applications. Given such uncertainty, most numerical models neglect the effect of supercooling entirely [13–15], which results in overly-optimistic predictions of system performance [16]. Thus, from a practical point of view for successful implementation of large- scale PCM systems, it is crucial to accurately predict their supercooling temperatures.

For applications with narrow temperature ranges (i.e buildings, refrigeration, medicine), supercooling can be even more limiting. As an example, consider a PCM with a transition temperature of 20 °C used in building applications. If the lowest temperature the building can reach is 15 °C, then the PCM cannot supercool more than 5 degrees. If it does, the PCM will never crystallize and the material will be rendered inert, so that the only thermal energy storage comes from sensible – as opposed to latent – heat. Lacking predictive power, industry often cannot rely on lab-scale supercooling data reported in the literature, and instead find it necessary to perform their own time-consuming and expensive large-scale testing before installations. Using lab-scale data to predict supercooling behavior for larger systems is the key bottleneck in translating research into application.

To overcome the bottleneck, this paper explores how nucleation theory is inherently coupled to thermal transport phenomena. We first establish a method to characterize the nucleation rate in PCMs using common lab-scale thermal analysis instrumentation (DSC, DTA, T-History). Using the nucleation rate from lab-scale, we show how to predict the supercooling temperature as a function of volume and cooling rate for isothermal (slow) cooling. We then generalize that analysis to calculate the supercooling temperature and transience for an arbitrary geometry, volume, and thermal boundary conditions which results in temperature gradient in the system– enabling accurate supercooling predictions for any system given lab-scale data. The analysis can be used in conjunction with existing numerical methods to accurately incorporate supercooling into phase change models, thus combining material modeling with system modeling. Finally, we validate the methodology outlined by comparing to experimental results in Neopentyl Glycol, which shows how the model successfully predicts the changes in subcooling temperature across a large range of cooling rates (2 orders of magnitude) and volumes (3 orders of magnitude).

1. **Determining the Nucleation Rate of a PCM**

Nucleation is a stochastic process and as such can be described with an appropriate statistical distribution. As mentioned, phenomenological models exist (e.g. Classical Nucleation theory [10]) to describe that distribution but are not useful for experimental characterization given that multiple input parameters are unknown for the systems of current interest. It is instead more convenient to describe nucleation as a non-homogeneous Poisson process [11,17–21]. A Poisson process is a purely statistical model for a series of discrete events where the average time between events is known, but the exact timing of events is random and memoryless. These assumptions are consonant with nucleation theory, where the average time between nucleation events is given by the reciprocal of the average nucleation rate, but the process is stochastic so that the exact timing of nucleation events is random. The Poisson process further assumes that the nucleation events are independent and cannot occur at the same time. Nucleation is said to be a "*non-homogeneous*" Poisson process because the rate parameter, which in this case is the nucleation rate, may not be constant over time (e.g. when the temperature changes with time). Note that this use of "non-homogenous" is completely different from the nucleation itself being homogeneous or heterogeneous.

The non-homogeneous Poisson distribution has been used in fundamental nucleation studies of metals [11,21], and in solution-crystallization processes [22,23], but not previously for characterizing PCMs. Here we tailor the application of the non-homogeneous Poisson distribution for the workflow of PCM characterization.

We begin by introducing the properties of the distribution of subcooling temperatures. The cumulative distribution function is described by $CDF(t) = 1 - e^{-\int_0^t \lambda(t') dt'}$, and the survivor function is $\chi(t) = 1 - CDF(t)$ where $\lambda(t)$ is the Poisson rate parameter, which corresponds to the system-wide nucleation events per second at time *t*. In general, $\lambda(t)$ depends on the number of nucleation sites at the surface and within the volume. Thus, $\lambda(t)$ is the sum of two contributions so that $\lambda(t) = VJ_V(t) + AJ_A(t)$ where V is the system volume, A is the system surface area, and $J_V(t)$ and $J_A(t)$ are the volume-specific and area-specific nucleation rates, respectively. In a given system, normally either $J_V(t)$ or $J_A(t)$ will dominate so the nucleation rate can be normalized by only the surface area or the volume, such that $J(t) = \frac{\lambda(t)}{V}$ or $\frac{\lambda(t)}{A}$. We note that $J_V(t)$ has contributions from both heterogeneous and homogeneous nucleation sites within the volume of the system, whereas $J_A(t)$ has contributions only from heterogeneous nucleation sites at the surface of the system. For the rest of this text we will assume that $VJ_V(t) \gg AJ_A(t)$ so that the system's nucleation physics scales only with the system volume. We also note that in principle, most modeling results obtained below could be easily translated to a system with surface area-dominated nucleation physics $(AJ_A(t) \gg VJ_V(t))$ by suitable exchanges $J_V(t) \leftrightarrow J_A(t)$ and $V \leftrightarrow A$.

In PCM characterization, the temperature is the independent variable of interest – not the time. Moreover, our goal is to couple nucleation with thermal transport, so we make a change of variables from $t$ to $T$ as the independent variable, such that $T(t) = T$, $\frac{dT}{dt} = \dot{T}$, and as the initial condition $T(t=0) = T_m$, where $T_m$ is the equilibrium phase transition temperature. In DSC, the most popular PCM characterization technique, the cooling rate is typically constant. We can then simplify by assuming $\dot{T} = -\beta$, where $\beta$ is the constant cooling rate in degrees per second, defined to be a positive quantity. If β is not constant, $\dot{T} = \dot{T}(t)$ in the integration. Applying this change of variables to the survivor function yields:

$$\chi(T) = e^{-\frac{V}{\beta} \int_{T_m}^{T} J(T') \, dT'} \tag{1}$$

The goal is to compute the volume-specific nucleation rate, $J_V(T)$, from experimental data, so we invert the survivor function, $\chi(T)$ [18]:

$$J_V(T) = -\frac{1}{V} \frac{\beta}{\chi(T)} \frac{d\chi(T)}{dT} \tag{2}$$

The distribution of supercooling temperatures, $\chi(T)$, can be determined for a given PCM sample from cooling experiments, and thus from Eq. (2) the nucleation rate can be calculated for the material given the sample volume and the experimental cooling rate. It is crucial that the sample volume be cooled uniformly during supercooling experiments, such that the Biot Number ($Bi = \frac{hL}{k}$) is less than 0.1, where $L$ is the length scale of the sample which can be estimated as $L = \frac{V}{A}$, $h$ is the convection coefficient, and $k$ is the thermal conductivity. An example of this $J_V(T)$ extraction procedure is discussed in Section 6. We emphasize that this procedure is also valid for surface-dominated nucleation, in which case Eq. 2 would be normalized by the surface area instead of the volume and the result is $J_A(T)$ rather than $J_V(T)$. To determine whether volume-based or surface-based nucleation dominates, the nucleation rate can be measured for multiple, different sized samples in a DSC pan and should be normalized by both the volume and the surface area. The volume-normalized and surface-normalized nucleation rates can then be plotted against supercooling temperature, and whichever normalization collapses on a single line with zero intercept is the dominant nucleation mechanism.

Equation 2 can be evaluated numerically, but in Sections 2-5 it will prove useful to have an analytical form of the nucleation rate. To that end, it is convenient to define a fitting function to the normalized nucleation rate [24]:

$$J_V(T) = \gamma \Delta T^n \tag{3}$$

where $\Delta T$ is the difference between the thermodynamic equilibrium phase change temperature and the actual temperature, i.e., the supercooling. Equation 3 captures the nucleation behavior of a material with just two empirical parameters, $\gamma$ and $n$. We argue that it is very important that researchers report $\gamma$ and $n$, or some other description of $J_V(T)$, when characterizing new PCMs. As will be shown, once $\gamma$ and $n$ are known, the supercooling behavior of the material in an arbitrary thermal and geometric system can be predicted.

## 2. Predicting supercooling for arbitrary thermal and geometric systems – General theory

Using the nucleation rate determined in Section 1, the goal is to determine the average time it will take for a PCM to nucleate given the system geometry, volume, material properties, and thermal boundary conditions.

In Section 1 we established that the volumetric nucleation rate can be described by just two parameters and the subcooling temperature, $J_V(T) = \gamma \Delta T^n$, characterized at the lab scale (e.g. DSC). At this scale ($V \approx 10 \ \mu L$), we can ensure that the temperature distribution within the PCM is approximately uniform by controlling the sample thickness ($L < 1 \ mm$) and the cooling rate ($\beta \approx 10 \frac{°C}{min}$). By doing this, the nucleation probability, $J_V(T)$, also becomes approximately uniform in space, ensuring accurate volumetric nucleation rates. In a general system at larger scale, however, the temperature of the material varies considerably with both position and time, and consequently, so does the nucleation probability. *The system's nucleation probability is therefore inherently coupled to thermal transport phenomena.*

Qualitatively, this indicates that the subcooling probability in each material element within the PCM is governed by its own statistical distribution, $i$, dependent only on the local subcooling history at that point, $T(\mathbf{x}_i, t)$. Each distribution is characterized by its Poisson rate parameter, $\lambda_i$ and resides over volume element $dV_i$. To determine the global nucleation probability of the material, the distributions must be combined into one, and the combined probability distribution of the sum of independent random variables is equal to the convolution of their individual distributions. Fortunately, for a non-homogeneous Poisson process, the convolution is simply the sum of the individual rate parameters [25]:

$$\sum_i^N Poisson(\lambda_i) = Poisson\left(\sum_i^N \lambda_i\right) \qquad (4)$$

where $N$ is the total number of independent distributions; here, $N$ is the number of discrete material elements considered. We can then define a global, or an effective, rate parameter, $\lambda_{eff}(t) = \sum_{i=1}^N \lambda_i(x,y,z,t)$, or $\lambda_{eff}(t) = \sum_{i=1}^N J_i(x,y,z,t) V_i(x,y,z)$. Passing into the limit of infinitesimally small volume elements, the effective rate parameter of the system at a given time can be rewritten as:

$$\lambda_{eff}(t,\gamma,n) = \iiint J(x,y,z,t) dx dy dz = \iiint \gamma (T(x,y,z,t) - T_m)^n dx dy dz \qquad (5)$$

and the CDF becomes

$$CDF(t,\gamma,n) = 1 - e^{-\int_0^t \lambda_{eff}(t') dt'}. \qquad (6)$$

The probability density function (PDF) is obtained directly from equation 6 as $PDF(t) = \frac{dCDF(t)}{dt}$. The average time that it takes for the first nucleation event to occur can be calculated as the first moment of the PDF, and the standard deviation as the second:

$$t_{avg}(\gamma, n) = \int_0^\infty t\, PDF(t)dt \qquad (7)$$

$$\sigma_t(\gamma, n) = \int_0^\infty t^2\, PDF(t)dt \qquad (8)$$

Thus, if $T(x, y, z, t)$ is known, by applying Eqs. 5-8 the average time until the first nucleation event of the system, and the standard deviation of the distribution of nucleation-onset times, can be calculated from only the two nucleation parameters determined in Section 1. We note that the parameterization of $J_v$ is not unique, and the same procedure (equations 5-8) can be followed with an arbitrary parametrization of $J_v$.

Analytical solutions for $T(x, y, z, t)$ are not available for all but the simplest geometries and boundary conditions, so in general this procedure must be carried out numerically. Equations 4-8 are naturally discretized in space (index *i*) and time, and can be easily incorporated into existing numerical schemes for PCMs such as finite element methods, the enthalpy method [14], effective heat capacity method [26], and the heat source method [27]. Using any of these methods, *T(x,y,z)* can be determined at each time step, and the integral of the nucleation rate as a function of *T(x,y,z)* over the volume in Eq. 5 can be calculated to determine the effective global nucleation rate at time *t*. Stepping through time, $\lambda_{eff}(t)$ can be calculated, and from $\lambda_{eff}(t)$, the CDF, PDF, and then average time to nucleation can be determined.

This output can then be used to identify when to trigger nucleation in subsequent simulations with single-phase initial conditions, thus providing high resolution and high-fidelity initial conditions for crystallization studies, or nucleation triggers for cyclic performance simulations of PCMs. Conversely, the probability of nucleation at each discrete element and time step can be calculated from equations 4-7, which makes this analysis suitable for Monte Carlo based nucleation simulations. Incorporating either workflow into numerical models will take into account the hysteresis inherent in PCM cycling caused by supercooling, and enabling more realistic calculations of PCM charge/discharge times -- both of which are critical in predicting PCM performance in real-world applications.

## 3. Approximate Solution – Uniform Temperature Distribution

Analytical solutions to equations 6-8 are intractable for most boundary conditions. However, for a spatially uniform i.e. lumped, temperature distribution such that $T(x, y, z, t) = T(t)$, and a constant cooling rate $\beta$, equation 5 is greatly simplified and the CDF can be evaluated analytically:

$$CDF(T) = 1 - e^{-\frac{V}{\beta} \int_{T_m}^{T} J(T) \, dT} = 1 - e^{-\frac{V}{\beta} \frac{\gamma (T-T_m)^{1+n}}{1+n}} \qquad (9)$$

Using this CDF, equations 6-8 can be calculated (with dependent variable T instead of t), and it has been shown [24,28,29] that the temperature at which nucleation will occur on average, i.e. the average supercooling temperature, as a function of volume and cooling rate can be expressed as:

$$\Delta T_{avg}(V, \beta) = \beta^{\frac{1}{n+1}} \left(\frac{n+1}{\gamma V}\right)^{\frac{1}{n+1}} \Gamma\left(\frac{n+2}{n+1}\right) \qquad (10a)$$

where $\Gamma$ is the gamma function.

The average time until nucleation is triggered is then $t_{avg}(V, \beta) = \frac{1}{\beta}\left(T_m - T_{avg}(V, \beta)\right)$:

$$t_{avg}[s] = \frac{1}{\beta}\left(\beta^{\frac{1}{n+1}} \left(\frac{n+1}{\gamma V}\right)^{\frac{1}{n+1}} \Gamma\left(\frac{n+2}{n+1}\right)\right). \qquad (10b)$$

When $\beta$ has its traditional units of [°C/min], this expression for $t_{avg}$ will be in units of [min]. It also is important that the same units are used for V in both Eq. (10b) and the definition of $J_V$ used in evaluating $\gamma$ from Eq. (3), whether [L] and [L], [m³] and [m³], etc.

We emphasize that for PCMs in large-scale applications, a lumped temperature distribution is generally a bad approximation. It is valid when $\frac{hL}{k} < 0.1$. For natural convection a typical value of $h \approx 10 \, \frac{W}{m^2 K}$ and for PCMs $k \approx 0.1 - 0.5 \, \frac{W}{m*K}$, so the thickness of the PCM must be less than ~1 - 5 mm to justify the lumped approximation, which is not generally useful for large-scale applications. For small-scale applications such as micro-encapsulation, $L \approx 1 \, \mu m$ so equation 10 is generally valid if volumetric nucleation dominates. We can use equation 10 then to interpret the well-known phenomenon that micro-encapsulated PCMs typically exhibit much larger supercooling than a macroscopic-volume sample. We see from equation 10a that $\Delta T_{avg}$ scales inversely with volume, specifically $\Delta T_{avg} \propto V^{-\left(\frac{1}{1+n}\right)}$, so for a smaller volume the supercooling increases, which has crippled many micro-encapsulation efforts [4,30]. This scaling occurs because a reduction in volume implies a reduction in the number of nucleation sites, and the probability of nucleation decreases with the decrease in the number of nucleation sites. Equation 10 can be used as a guide, therefore, in determining promising candidates for micro-encapsulation applications. It is important to note, however, that the volume dominated nucleation may be eclipsed by surface nucleation sites as the length scale decreases. In fact, this has been used to combat supercooling problems in microencapsulated PCMs. To do this, researchers choose micro-encapsulation materials that have lower surface energies with the PCM, promoting surface nucleation [31,32]. Equation 10a can then be seen as a worst-case scenario supercooling temperature for micro-encapsulation. If the nucleation does not become surface dominated by whatever means, then the supercooling temperature predicted by equation 10a is the lowest average supercooling temperature expected. Any surface effects would serve to decrease the extent of supercooling.

## 4. A Standardized Definition for the Supercooling Temperature

The supercooling temperature is fundamentally different than the phase transition temperature. The phase transition temperature is described by information embedded in the atomic details of the system and are well-described by deterministic equilibrium thermodynamics. The supercooling temperature, on the other hand, is *not* a fundamental property of the material because it depends on the kinetics. Phenomenologically, the kinetics are dictated by the distribution of nucleation sites and the energy barriers associated with those sites relative to the thermal energy scale. The number of nucleation sites scales with the volume and/or surface area of the material, and the rate at which those nucleation sites gain access to the available thermal energy from the environment is strongly material and problem dependent. The volume/surface-area scaling, coupled to a strong material-transport dependence, makes defining a meaningful supercooling temperature for a given material challenging. To date, many researchers report supercooling values observed in DSC/DTA, but this value provides only a single point in a complex space defined by equations 6-8. For this reason, researchers testing the same material under different experimental conditions often observe and report different supercooling temperatures.

To make more meaningful comparisons, it would be helpful for the thermal energy storage community to agree upon a standardized reference system for which a supercooling temperature can be defined. For our work, we choose a reference system that has a volume of 1L with cooling rate $\beta = 1 \frac{°C}{min}$ as a demonstration. Treating the system as lumped, we use the solution for a uniform temperature distribution (equation 10a), to obtain an expression for the supercooling temperature of this reference system:

$$\Delta T_{Supercooling,std} = \left(\frac{1}{60}\left[\frac{°C}{s}\right]\right)^{\frac{1}{n+1}} \left(\frac{n+1}{\gamma \cdot 10^{-3}[m^3]}\right)^{\frac{1}{n+1}} \Gamma\left(\frac{n+2}{n+1}\right) \quad (11)$$

*To be clear, this definition says that a 1L volume of PCM cooled uniformly at* $1\frac{°C}{min}$ *will nucleate at* $\Delta T_{Supercooling,std}$. With this standardization of β and V, this definition of $\Delta T_{Supercooling,std}$ depends only on $\gamma$ and $n$. Researchers testing the same material under different experimental conditions will still observe different supercooling temperatures, but by determining the nucleation parameters $n$ and $\gamma$ they can now agree on the supercooling temperature of this reference system. The supercooling performance of a PCM can then be compared by comparing the supercooling behavior of each PCM in this reference system.

## 5. Experimental Validation

To validate the methodology outlined in Sections 1-5, we characterize the nucleation rate of neopentyl glycol's (NPG) solid-solid phase change using DSC and then perform larger-scale cooling experiments to test the predictions of Sections 3 and 4. We choose NPG because its transition is sharp, near room temperature ($T_m = 40.8\ °C$), and it doesn't interact with the aluminum DSC pans so volumetric nucleation dominates. NPG was acquired from Sigma Aldrich at 99% purity, and TA Instrument's DSC 2500 with indium temperature calibration was used.

10 mg of NPG was cycled 150 times from $25°C$ to $50°C$ in a DSC at a heating and cooling rate of $10\frac{°C}{min}$. The cooling curves from all 150 runs are shown in figure 1a. The supercooling temperatures were determined as the first deviation from the linear baseline heat flow signal ($\frac{W}{g}$) that exceeded the minimum accuracy of heat flow on the DSC (first detectable deviation). For the DSC 2500, the minimum deviation is 20 $\mu W$. There was no sign of aging (see SI Section 1). This set of 150 supercooling temperatures was then used to calculate the survivor function by taking the ratio of non-nucleated (surviving) samples at a given temperature and dividing by the total number of samples. For temperatures greater than $29.3°C$, zero samples out of the 150 had nucleated, so the survivor function is equal to 1 (all samples survived) for T>29.3°C. Similarly, for temperatures below $27.1°C$, all samples have nucleated, so the survivor function is equal to zero (no samples survived) for T<27.1°C, as shown in figure 1b. From the survivor function, the nucleation rate as a function of temperature was calculated using equation 2 and fitted to the power law, $J_v(T) = \gamma \Delta T^n$, resulting in $\gamma = 4.2 * 10^{-33} \frac{1}{K^n s m^3}$ and $n = 35.87$ as shown in figure 1d. To show the goodness of fit, the survivor function (equation 1) is plotted using $J_v(T) = 4.2 * 10^{-33} \Delta T^{35.87}$ in figure 1b. The smooth line shows the survivor function given by the fitted analytical nucleation rate. There is good agreement between the fit and the data. Using equation 11, the standardized supercooling temperature for the reference system defined in Section 5 for NPG is then $\Delta T_{Supercooling,std}=8.85°C$ below the equilibrium transition temperature of $40.8°C$.

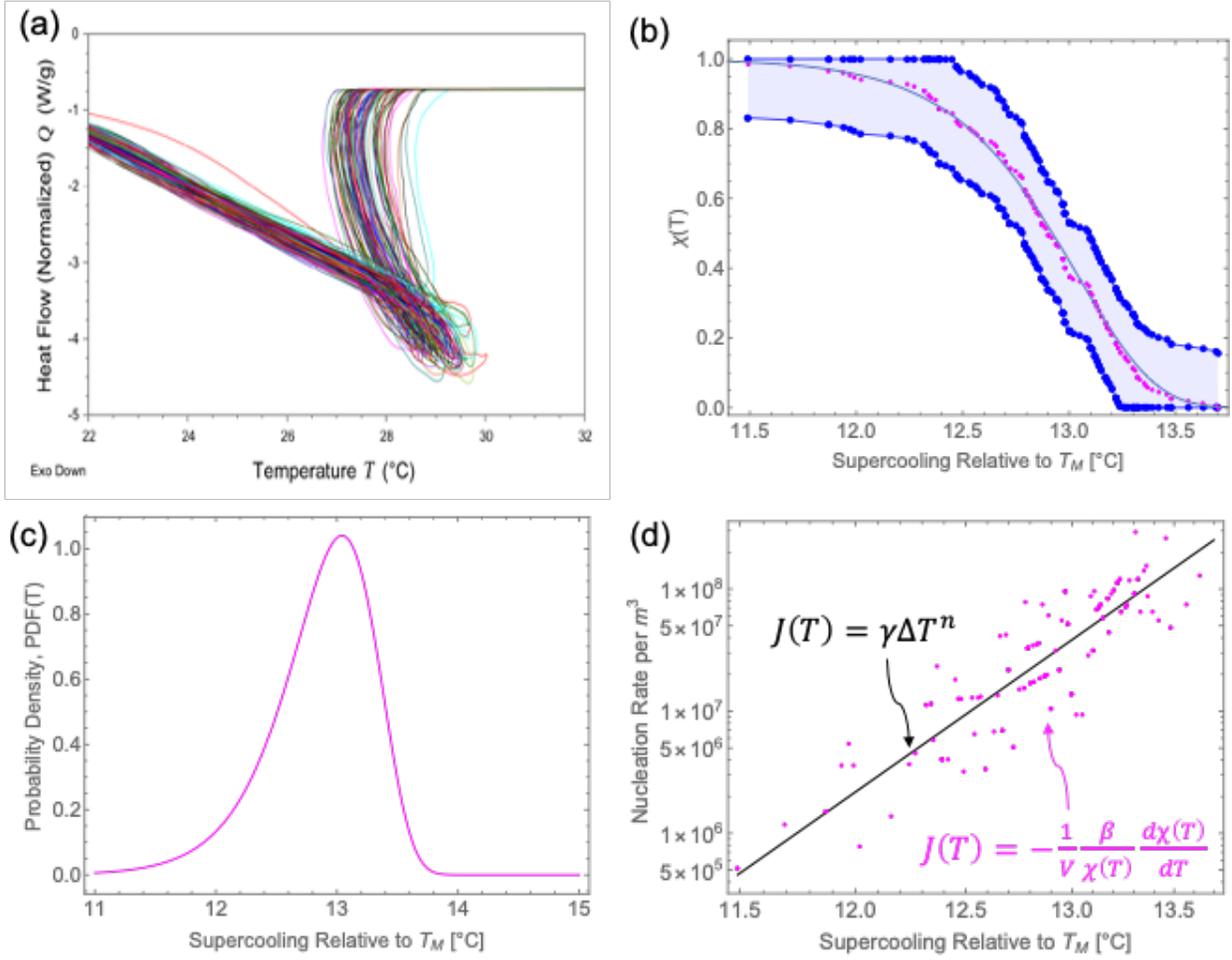

Fig. 1: (a) 2 Raw DSC data for 150 cooling cycles of 10 mg of NPG at β=10 $\frac{°C}{min}$. Each cooling curve exhibits a phenomenon known as recalescence, in which immediately after the onset of a phase transition the rate of heat released is temporarily greater than the rate of cooling, so the material heats itself up for a short time. From these curves, the distribution of supercooling temperatures is recorded. (b) shows the survivor function (magenta) discussed in Section 1, which was calculated from the distribution of supercooling temperatures given by the DSC data in (a). A 95% confidence interval using the DKW inequality is shown in blue surrounding the empirical survivor function. The solid black line shows the fitted CDF from the nucleation parameters, $\gamma$ and $n$ using equation 9. (c) The PDF computed from the survivor function, $PDF(t) = \frac{dCDF(t)}{dt}$. (d) shows the pointwise nucleation rate (magenta) calculated using a two-point finite difference form of equation 2, and its fit (black solid line) using equation 3. Panels (b)-(d) are referred to $T_m = 40.8\ °C$.

*Uniform Temperature Distribution Approximation*

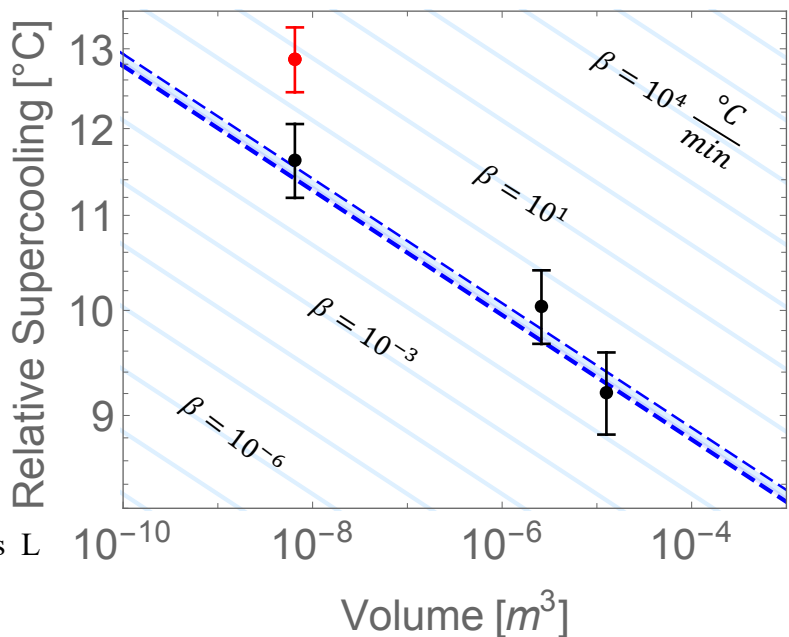

Using the nucleation parameters characterized via DSC, we can predict how the supercooling temperature will change with volume and cooling rate from equation 10a. To test this prediction, we run cooling experiments for NPG at different volumes spanning 3 orders of magnitude ($V = 6.0 * 10^{-9}, 2.6 * 10^{-6}, 1.3 * 10^{-5} \, m^3$). For these experiments, the appropriate volume of NPG was melted in an oven and poured into aluminum weigh-boats such that the thickness L was uniform and less than $0.5 \, mm$, ensuring a uniform temperature distribution during cooling (see SI Section 2). Three thermocouples with uncertainty $\pm 0.1°C$ were taped to the opposite sides of each aluminum pan, and the samples were cycled 20 times in an oven at a cooling rate of β=$0.1 \frac{°C}{min}$. Nucleation temperatures were recorded as the average supercooling value from each thermocouple for a given sample.

Fig. 2: Experimental vs predicted supercooling temperatures for NPG at different volumes and cooling rates for the uniform temperature distribution approximation. The black dots represent the experimental averages and the vertical lines their standard deviations, all for an experimental cooling rate of β=$0.1 \frac{°C}{min}$. The shaded blue region and the dashed blue lines represent the average predicted supercooling temperatures from equation 10a for a cooling rate of $0.1 \frac{°C}{min}$, bounded by one predicted standard deviation, using $\gamma = 4.2 * 10^{-33}$ and $n = 35.87$ determined from Fig. 2, The red point is another experiment using a faster cooling rate of $10^1 \frac{°C}{min}$.

We note that the crystal growth rate in NPG appears to be very fast, so that all thermocouples record the nucleation event within 5 seconds of each other and thus there is close agreement among the measurements. The experimental results are plotted against the predictions made from the nucleation parameters in Figure 2 Variations of the prediction with $\beta$ are shown, and it can be seen that changing $\beta$ simply offsets the curve, but not the slope, in this log-log plot. Experimental vs predicted values for constant volume but varying $\beta$ are shown in figure 2 for $\beta = 10^{-1} \frac{°C}{s}$ and $\beta = 10^1 \frac{°C}{s}$, and the predicted values agree closely with experimental results.

*General Theory*

To test the predictions in Section 4 we keep the volume constant at $V = 2.6 * 10^{-6} \ m^3$ but vary the aspect ratio, so that the transient cooling is no longer lumped. By varying the geometry, we change the local thermal conditions at each nucleation site. To do this, we iteratively melted $2.6 * 10^{-6} \ m^3$ of NPG in an oven and cast them into PTFE tubes of varying diameter, ranging from 0.635 to 1.588 cm. We note that the PTFE was chosen because the tubes do not interact with NPG, and they are non-stick for easy removal. Once removed from the PTFE forms, the NPG cylinder (shown in blue in Fig 3.) was bridged between two columns of foam (diagonal black hatched lines) such that its end faces were insulated, while its circumferential surface was open to the environment (i.e convective boundary conditions).

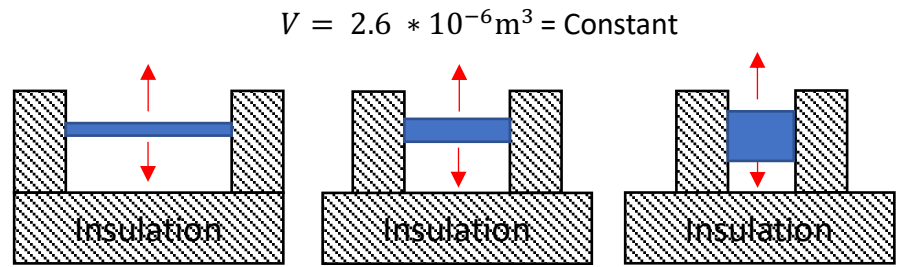

Fig 3. Experimental setup to test general theory. NPG (blue) was cast into cylinders of varying lengths and diameters, but constant volume. The cylinders were suspended between two foam insulation columns such that the end faces were insulated. Heat transfer is in the radial direction (red arrows) during these cooling experiments.

Thermocouples were placed behind aluminum foil contacting one face of the cylinder, so that the thermocouple tip had no direct contact with the NPG. The NPG cylinders were then equilibrated at $T_o = 50°C$ in an oven for several hours. Then each cylinder was transferred to an environment held at T$_\infty$=25°$C$ to cool by free convection, and the time until nucleation was recorded for each aspect ratio and compared against the predictions given by equations 6-8. For these predictions, edge effects were important so we used COMSOL Multiphysics to calculate $T(x, y, z, t)$. To make these predictions, we calibrated for the free convection coefficient by placing a thermocouple on the outer surface of an NPG cylinder and measuring the temperature vs time curve and fitting numerical solutions for temperature vs time curves to that data to determine $h$ (see SI 2). This calibration yielded $h = 18 \frac{W}{m^2 K}$, taken as constant, which is a reasonable value for free convection and was used for the predictions in figure 4.

The predictions of equations 6-8 are plotted (solid blue line) against experimentally determined nucleation times (black dots) in Figure 4. The shaded blue region denotes predictions that are one standard deviation above and below the average nucleation time. There is excellent agreement between equations 6-8 and experimental values. We include two additional curves on the figure to contextualize the importance of including temperature gradients and convective boundary conditions, as described next.

Previous related work [24,28,29] had two key simplifying assumptions: (i) a lumped temperature response (that is, $T(t)$ only); (ii) constant cooling rate, $\beta$. The present work focuses on relaxing both of these requirements by considering non-lumped temperature response (that is, $(T(x, y, z, t))$) and convectively coupled cooling which leads to variable cooling rate. To demonstrate the effect of relaxing these requirements, we show two additional predictions in Figure 4. First, we compare to the lumped constant cooling rate case, where the average nucleation time is given by equation 10b Because the physical system experienced convective cooling and not constant cooling rate, we approximated a constant cooling rate, $\beta$, by evaluating the convective cooling rate at $t = 0$ from Newton's Law of Cooling. This gives $\frac{dT}{dt}|_{t=0} = -\beta = -\frac{hA}{mC_p}(T_o - T_\infty)$. We used this $\beta$ and the system volume, $V = 2.6 * 10^{-6} m^3$, as inputs into equation 10b and the results are plotted as the gray dotted line in figure 4. It can be seen that using the lumped approximation with constant cooling rate gives average nucleation times that are ~1.5x lower than that given by the more detailed treatment of the non-lumped $(T(x, y, z, t))$ with equations 5-8.

Next, we maintained the lumped approximation and relax the constant cooling rate assumption by using Newton's law of cooling $T(t) = T_\infty + (T_o - T_\infty)e^{-\left(\frac{hA}{mC_p}\right)t}$ to describe the time-dependent convective cooling, which was used as the input into equations 6-8. This ensures that the sample is convectively-coupled to the surroundings with the same value of $h = 18 \frac{W}{m^2 K}$ used in the more detailed simulations. The predictions for this lumped, convective cooling are shown as the gray dashed line in figure 4, and now overpredict the actual cooling time by around a factor of 2. It overpredicts because the lumped assumption ignores large temperature gradients which generally arise near the material surface. The large temperature gradients lead to much lower temperature near the surface, which catalyzes nucleation. Because the lumped assumption ignores these temperature gradients, the catalyzed effect on nucleation is missed, leading to over-prediction. We show these comparisons to highlight the fact that the lumped assumption can easily cause large (~2x) errors. Thus, accounting for the temperature gradients within the sample during cooling is very important for an accurate description of a sample's supercooling behavior.

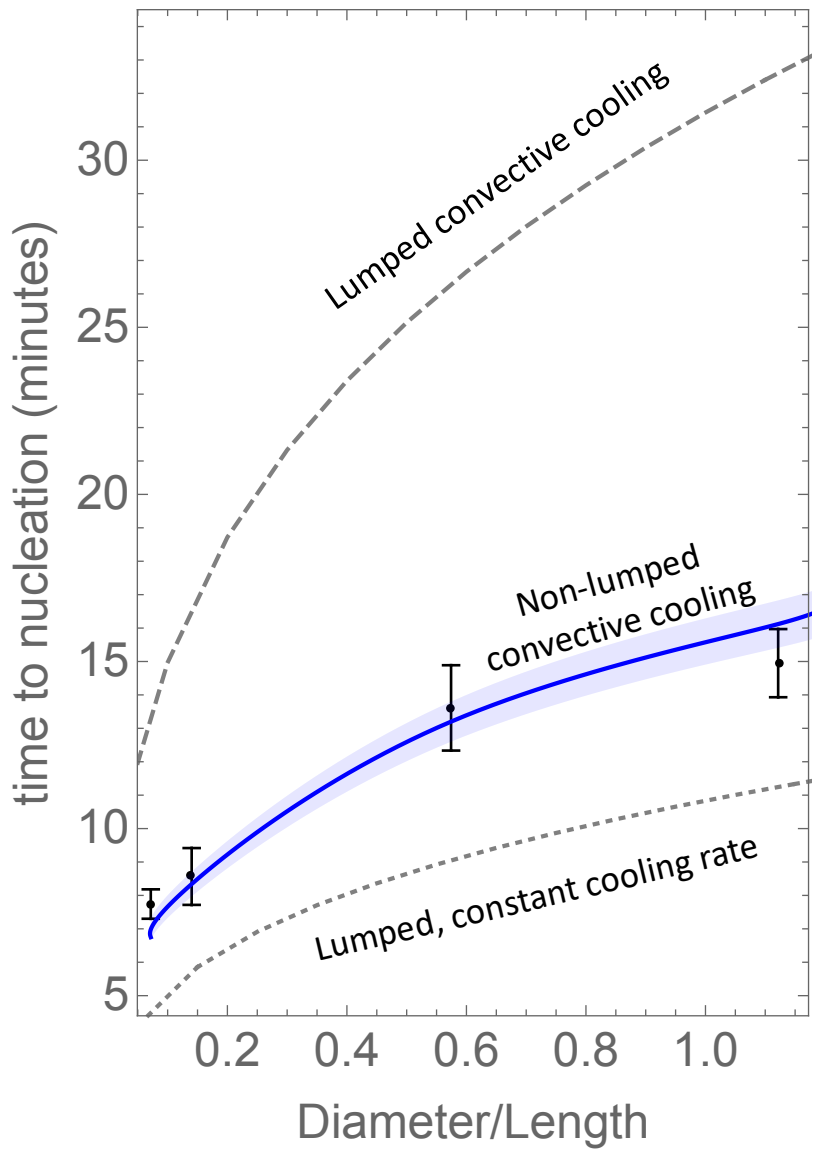

FIG 4: Experimental vs Predicted supercooling times for NPG at fixed volume with varying aspect ratio (see figure 3) The black dots represent the experimental averages and their standard deviations on the vertical. The solid blue line represents the predicted supercooling times from equations 5-8 using $\gamma = 4.2 * 10^{-33}$ and $n = 35.87$. The blue shaded zone represents $\pm 1$ standard deviation from equation 8. The gray dashed and dotted lines show predictions for lumped convective and lumped constant cooling cases, and are included to highlight the importance of taking into account temperature gradients (non-lumped) to correctly predict the supercooling behavior of a system.

**Conclusion**

Using lab scale experimental data to predict supercooling performance in large scale thermal energy storage applications is crucial for the analysis and prediction of PCM performance metrics. This paper has outlined experimental characterization techniques for supercooling in thermal energy storage applications and developed a theoretical framework to use that characterization for prediction of supercooling in a generalized system. The analysis can be used in conjunction with existing numerical methods to accurately incorporate supercooling into phase change models, thus combining material modeling with system modeling. This framework has been validated by comparing to experimental results in neopentyl glycol, which shows how the model successfully predicts the changes in subcooling temperature across a large range of cooling rates (2 orders of magnitude) and volumes (3 orders of magnitude). To expand this framework, future efforts should explore the characterization of more exotic and complex materials (e.g. polymers, mixtures).

**Acknowledgement:**
This work was supported by Energy Efficiency and Renewable Energy, Building Technologies Program, of the U.S. Department of Energy under Contract No. DEAC02-05CH11231.

Supplementary Information

## 1. Experimental Supercooling Data and Quality of Distribution

10mg of NPG was cycled 150 times from $25°C$ to $50°C$ in a DSC at a heating and cooling rate of $10 \frac{°C}{min}$. The cooling curves from each run are shown in figure 1 of the main text. The supercooling temperatures were determined as the first deviation from the heat capacity that exceeded the minimum accuracy of heat flow on the DSC (first detectable deviation). For the DSC 2500, the minimum deviation is $20 \ \mu W$. The extracted values are plotted below as a function of trial number.

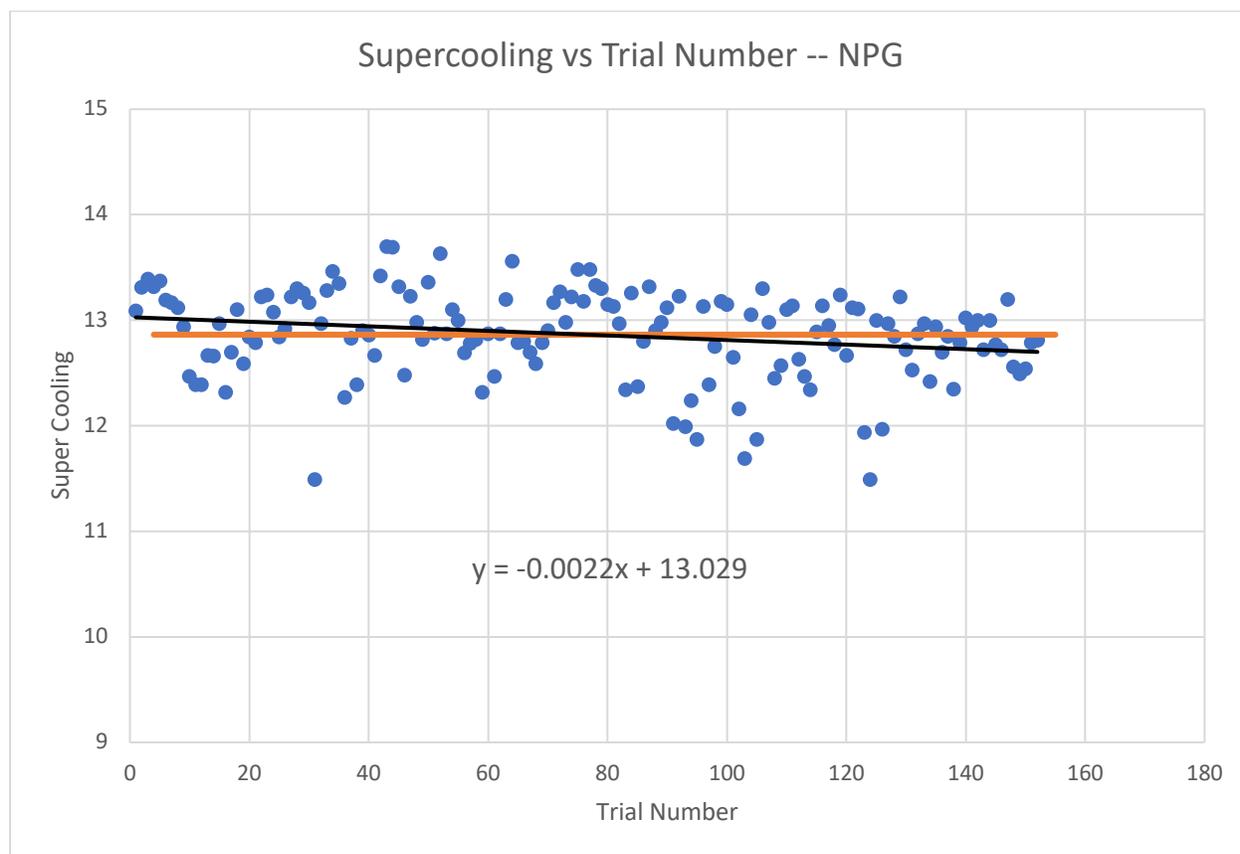

To check for signs of material aging, we provide a linear fit (black solid line) to the data, and compare that with the average of the distribution (orange solid line). It is seen that the average fit is nearly constant vs trial number, indicating little to no aging. In addition, we take the raw data and generate a normal probability plot. This plots the CDF of the experimental data vs the CDF of a theoretical normal distribution. If the experimental data is perfectly normally distributed about the mean, the plot would show a straight 45° line. As can be seen from the normal probability plot below, the experimental data is approximated well by a normal distribution, indicating that there's no aging or systematic bias.

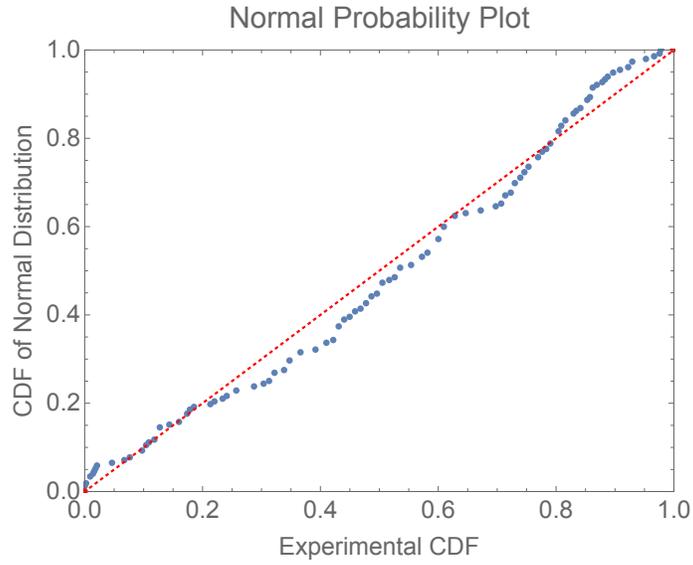

## 2. Experimental Setup for Uniform Temperature Distribution Approximation

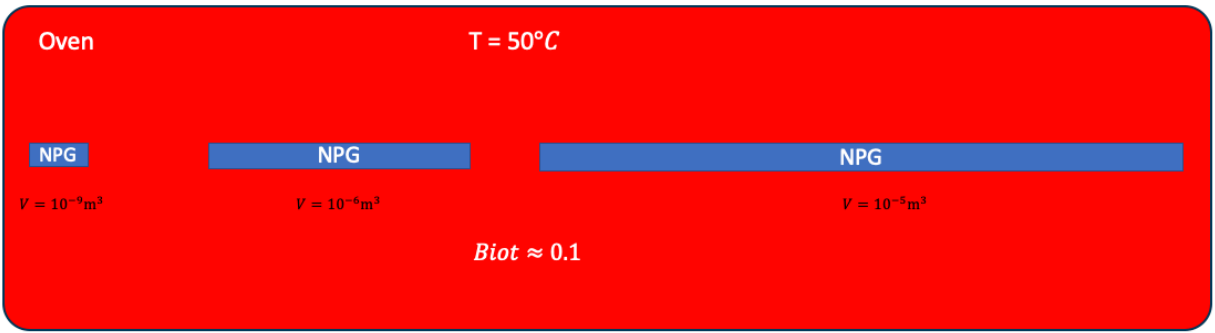

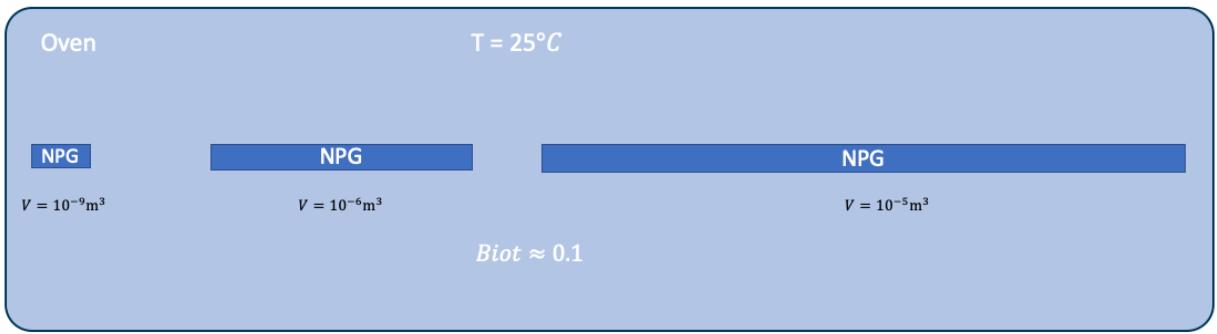

Cycled 20x

## 3. Convection Heat Transfer Coefficient Calibration

To make the predictions in figure 4, we needed to calibrate for the convection coefficient. We calibrated by placing a thermocouple on the outer surface of a $\frac{3}{8}$" diameter NPG cylinder, equilibrating the NPG at 50°C, and then measuring the temperature vs time curve after it was brought to ambient at $24°C$. To find the convection heat transfer coefficient associated with the experiment, we solved for the temperature vs time of the equivalent physical and geometric system in COMSOL Multiphysics, and fitted the numerical solutions to the experimental data to determine $h$. The calibration yielded $h = 18\frac{W}{m^2 K}$ which is a reasonable value for free convection and was used for the predictions in figure 4.

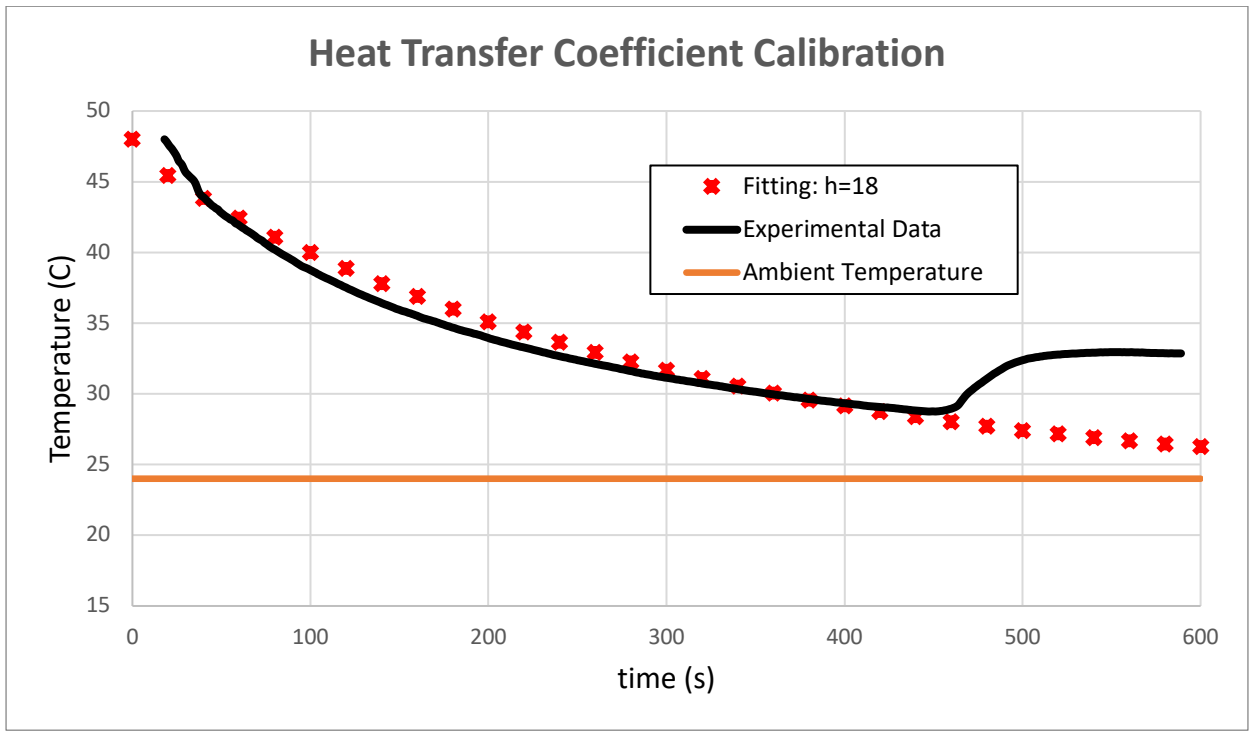

20